\documentclass[twocolumn,showpacs,amsfonts]{revtex4-1}
\usepackage{amsmath}
\usepackage{latexsym}
\usepackage{float}
 \usepackage{amssymb}
\usepackage{graphicx}
\usepackage{textcomp}
\usepackage{hyperref}
\usepackage{geometry}

\def\ba{\begin{eqnarray}}
\def\ea{\end{eqnarray}}
\def\be{\begin{equation}}
\def\ee{\end{equation}}
\def\bm{\begin{math}}
\def\me{\end{math}}

\newcommand{\dummy}

\begin{document}

\title{Kinetics of Ferromagnetic Ordering in 3D Ising Model: Do we understand the case of zero temperature quench?}
\vskip 0.5cm
\author{ Subir K. Das$^*$ and Saikat Chakraborty}
\affiliation{Theoretical Sciences Unit, Jawaharlal Nehru Centre for Advanced Scientific Research,
 Jakkur P.O, Bangalore 560064, India}

\date{\today}
\begin{abstract}
We study phase ordering dynamics in the three-dimensional nearest-neighbor Ising model, following rapid quenches from
infinite to zero temperature. Results on various aspects, viz., domain growth,
persistence, aging and pattern, have been obtained via the Glauber Monte Carlo simulations of the model on
simple cubic lattice. These are analyzed via state-of-the-art methods, including the finite-size scaling,
and compared with those for quenches to a temperature above the roughening transition. 
Each of these properties exhibit remarkably different behavior at the above mentioned final temperatures.
Such a temperature dependence is absent in the two-dimensional case for which there is no roughening transition.
\end{abstract}

\maketitle
\section{Introduction}
When a paramagnetic system is quenched inside the 
ferromagnetic region, by a change of the temperature from $T_i$ ($>T_c$) to $T_f$ ($<T_c$),  
$T_c$ being the critical temperature, it becomes unstable to fluctuations \cite{onuki,bray_phase,puri}. 
Such an out-of-equilibrium system
moves towards the new equilibrium via the formation and 
growth of domains \cite{onuki,bray_phase,puri}. These domains are rich in atomic magnets aligned in the same 
direction and grow with time ($t$) via the curvature driven motion of the interfaces \cite{bray_phase, %
Allen_cahn,puri}.
The interface velocity scales with
$\ell$, the average domain size, as \cite{bray_phase,Allen_cahn}

\begin{equation}\label{eq1}
\frac{d\ell}{dt} \sim \frac{1}{\ell}.
\end{equation}
This provides a power-law growth \cite{bray_phase,Allen_cahn}
\begin{equation}\label{eq2}
 \ell \sim t^\alpha,
\end{equation}
with $\alpha=1/2$. Depending upon the order-parameter symmetry and system dimensionality ($d$),
there may exist corrections to this growth law \cite{bray_phase}.
\par
Apart from the above mentioned change in the characteristic length scale, the domain patterns at different times, 
during the growth process, are statistically self-similar \cite{bray_phase,puri}. 
This is reflected in the scaling property \cite{bray_phase},
\begin{equation}\label{eq3}
 C(r,t) \equiv \tilde C(r/\ell),
\end{equation}
of the two-point equal-time correlation function $C$, where $r$ ($= |{\vec{r}_1-\vec{r}_2}|$) is 
the scalar distance between two space points and
$\tilde{C}$ is a master function, independent of time. A more general correlation
function involves two space points and two times, and is defined as \cite{puri}
\begin{eqnarray}\label{eq4}
 C(\vec{r}_1,t_{w};\vec{r}_2,t)=~~~~~~~~~~~~~~~~~\nonumber\\
\langle \psi(\vec{r}_1,t_{w}) \psi(\vec{r}_2,t)\rangle- 
\langle\psi(\vec{r}_1,t_{w})\rangle \langle\psi(\vec{r}_2,t)\rangle,
\end{eqnarray}
where $\psi$ is a space and time dependent order parameter. The total value of the order parameter,
obtained by integrating $\psi$ over the whole system, is not time invariant 
for  a ferromagnetic ordering \cite{bray_phase}. Thus, the coarseing in
this case belongs to the category of ``nonconserved'' order parameter dynamics \cite{bray_phase}.
For $\vec{r}_1=\vec{r}_2$, the definition in Eq. (\ref{eq4}) corresponds to the two-time autocorrelation 
function, frequently used for the study of aging property \cite{puri,fisher_huse,cor_lippi}  
of an out-of-equilibrium system, $t_w (\le t)$ being referred to as the waiting time or the age of the system. 
For the two point equal-time case, on the other hand,
$t_w=t$. The autocorrelation will henceforth 
be denoted as $C_{\rm ag}(t,t_{w})$. This quantity usually scales as \cite{puri,fisher_huse,cor_lippi,liu,satya_huse,yrd,henkel}
\begin{equation}\label{eq5}
 C_{\rm ag}(t,t_{w})=\tilde{C}_{\rm ag}(\ell/\ell_w),
\end{equation}
where $\tilde{C}_{\rm ag}$ is another master function,
independent of $t_w$, and $\ell_w$ is the value of $\ell$ at $t_w$. 
Another interesting quantity, in the context of phase ordering dynamics, 
is the persistence probability $P$ \cite{bray_per,satya_sire,satya_bray,derrida,%
stauffer,manoj_1,godreche,stauffer_jpa}. This is 
defined as the fraction of unaffected atomic magnets (or spins) and decays in a power-law fashion with time as 
\cite{bray_per} 
\begin{equation}\label{eq6}
P \sim t^{-\theta}.
\end{equation}
In the area of nonequilibrium statistical physics,
there has been immense interest 
in estimating the exponents $\alpha$ and $\theta$, as well as in obtaining the functional forms of $\tilde{C}$ and 
$\tilde{C}_{\rm ag}$, via analytical theories and computer simulations \cite{bray_phase,puri,bray_per}.
\par
In this work, we study all these properties for the nonconserved 
coarsening dynamics in the Ising model \cite{d_landau},
\begin{equation}\label{eq7}
H=-J\sum_{<ij>} S_{i} S_{j}, ~ J > 0, ~S_{i}=\pm 1,
\end{equation}
via Monte Carlo (MC) simulations \cite{d_landau}. 
We focus on $d=3$ and study ordering at $T_f =0$, for rapid quenches from $T_i=\infty$.
This dimension, particularly for $T_f=0$, received less attention compared to the $d=2$ case.
In $d=2$, the MC results for $C(r,t)$ are found 
to be in nice agreement with the Ohta-Jasnow-Kawasaki (OJK) function \cite{bray_phase,ojk,puri}
($D$ being a diffusion constant)
\begin{equation}\label{eq8}
 C(r,t)=\frac{2}{\pi}\sin^{-1}\bigg[ \exp\bigg(-\frac{r^2}{8Dt}\bigg) \bigg].
\end{equation}
This expression also implies $\alpha=1/2$, validity of which has been 
separately checked \cite{bray_phase,puri}. For the latter dimension, 
in the long time limit, the autocorrelation is understood to 
scale with $x$ ($=\ell/\ell_w$) as \cite{fisher_huse,liu,henkel,jia_suman}
\begin{equation}\label{eq9}
C_{\rm ag}(t,t_w) \sim x^{-\lambda},
\end{equation}
with $\lambda$ following a lower bound,
\begin{equation}\label{eq10}
 \lambda \geqslant \frac{d}{2},
\end{equation}
predicted by Fisher and Huse (FH) \cite{fisher_huse}.
In this case, also the persistence exponent $\theta$ has been 
estimated  \cite{stauffer,manoj_1,blanchard,saikat} for quenches  
to $T_f=0$. Furthermore, a few of these aspects, viz.,
the equal-time correlation function and domain growth, are understood to be independent of the value of $T_f$.

\par
Our interest in the zero temperature quench for $d=3$ was drawn by a 
controversy \cite{amar_family,holzer,cueille_jpa,corberi}, following
reports from computer simulations that observed 
$\alpha=1/3$ for this particular case, while the 
theoretical prediction is independent of $d$ and $T_f$. 
While addressing 
this issue, via simulations with large system sizes over long period, we made 
further interesting observations in other quantities. In this paper, 
we present these results on pattern, growth, aging and persistence. 
While our studies for pattern and aging are new, the results for growth and persistence 
are presented in forms different from a previous work \cite{saikat_pre}, to bring completeness to the discussion.
It appears, previous conclusion on the value of $\alpha$ was led by the presence
of an exceptionally long transient period, which was later hinted in Ref. \cite{corberi}.
In the ``true'' long time limit, the growth exponent is indeed $1/2$. Such a
trend we observe in the decay of the persistence probability as well.
On the other hand, the pattern and aging properties do not exhibit any crossover.
These results are very much different from those obtained for quenches to a temperature
above the roughening temperature.
Such temperature dependence does not exist in the $d=2$ case. 
Wherever necessary we presented results from the latter dimension as well.

\par
The rest of the paper is organized as follows. In section \textrm{II} we describe the methods.
Results are presented in section \textrm{III}. Finally, we conclude the paper in section \textrm{IV} 
by presenting a summary.

\section{Methods}
All our results were obtained via MC simulations of the Ising model  
using Glauber kinetics \cite{d_landau,glauber}, where, in each trial 
move the sign of a randomly chosen spin was changed. For $T_f=0$, the move
was accepted if it brought a negative change in the energy. On the other hand,
for $T_f>0$, whenever a trial move brought a higher energy contribution, the acceptance was decided by comparing
the corresponding Boltzmann factor with a random number (drawn from an uniform distribution),
a standard practice followed in the Metropolis algorithm \cite{d_landau}.

\par
In $d=2$ we have used square lattice and for $d=3$ the results are from simple cubic lattice.
All simulations were performed in periodic boxes of volume
$V=L^d$, $L$ being the linear dimension of a box, in units of the
lattice constant. For this model, the $d$-dependent  
critical temperatures \cite{d_landau} are $T_c \simeq 2.269 J/k_B$ ($d=2$) and 
$T_c \simeq 4.51 J/k_B$ ($d=3$), $k_B$ being the Boltzmann 
constant. Time in our simulations was measured in units of MC step (MCS), 
each MCS consisting of $L^d$ trial moves \cite{d_landau}. 
Unless otherwise mentioned, all results are presented after averaging 
over at least $10$ independent initial configurations, with $L=512$. 
For the rest of the paper we set $k_B$, tha lattice constant and the interaction strength ($J$) to unity.

\par
The average domain size was calculated in two different ways: (\textrm{i}) from 
the first moment of the domain size distribution $P_{d}(\ell_{d},t)$ as \cite{suman}
\begin{equation}\label{eq11}
 \ell=\int{\ell_{d}P_{d}(\ell_{d},t)d\ell_{d}},
\end{equation}
$\ell_d$ being the distance between two successive interfaces along any direction,
and (\textrm{ii}) using the scaling property of the correlation function as \cite{bray_phase}
\begin{equation}\label{eq12}
 C(\ell,t)=0.1.
\end{equation}
Unless otherwise mentioned, presented results are from Eq. (\ref{eq11}). For this
purpose, we have eliminated the noise in the configurations at nonzero temperatures, by
applying a majority spin rule \cite{suman}.
Note that the order-parameter $\psi$ here is equivalent to the Ising spin variable $S_i$. 
Thus, further discussions on the calculation of the other quantities are not needed.

\section{Results}

\begin{figure}[htb]
\includegraphics*[width=0.45\textwidth]{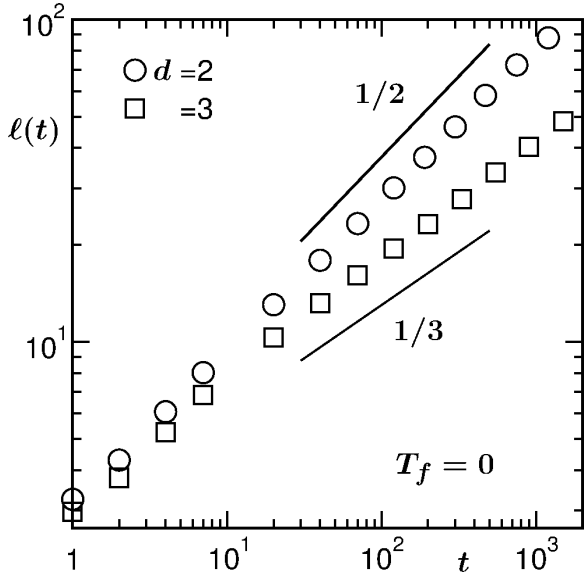}
\caption{\label{fig1}
Log-log plots of the average domain length, $\ell(t)$,  
vs time, for $T_f=0$. Results from both $d=2$
and $3$ are presented. In both the cases linear dimension of the system is $L = 200$. The 
solid lines correspond to two different power-law
growths, exponents being  mentioned in the figure.
}
\end{figure}

\begin{figure}
\includegraphics[width=0.35\textwidth]{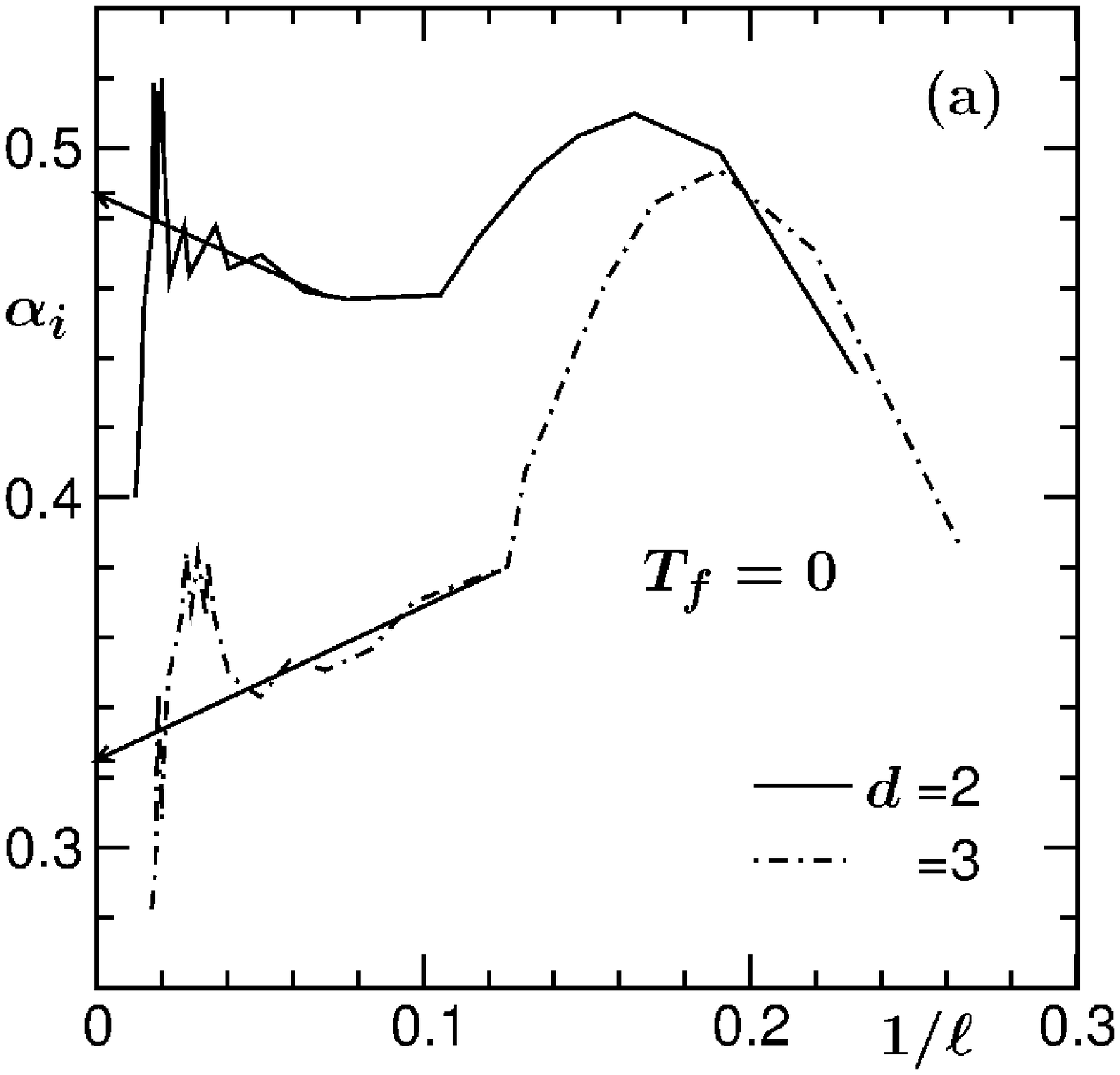} 
\hspace{3cm}               
\includegraphics[width=0.37\textwidth]{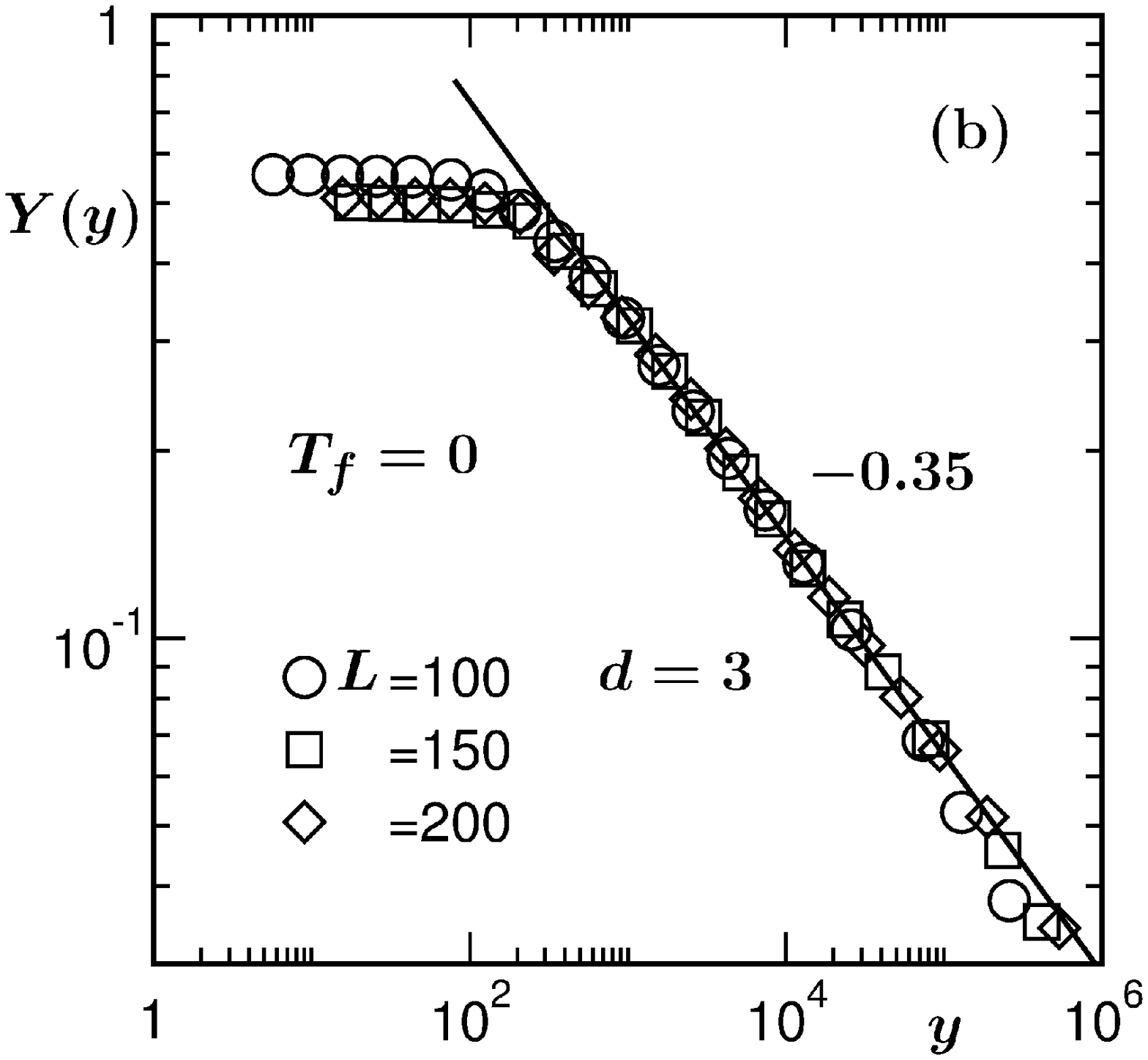}
\caption{\label{fig2}(a) Plots of the instantaneous exponent, $\alpha_i$, vs $1/\ell$, obtained 
from the data in Fig. \ref{fig1}.
(b) Finite-size scaling exercise for the $d=3$ results for $\ell(t)$. Here we have shown the scaling
function $Y$ with the variation of the dimensionless quantity $y$. $Y$ was obtained from the best collapse
of data from three different system sizes (mentioned in the figure).
The solid line corresponds to a power law decay with exponent $0.35$.
All results are from $T_f=0$.
}
\end{figure}

We start by showing the plots of $\ell$ vs $t$, for $d=2$ and $3$, 
at $T_f=0$, on a log-log scale, in Fig. \ref{fig1}. The system size considered here is comparable to
the early studies \cite{amar_family,holzer} in $d=3$.
The data for $d=2$ is 
clearly consistent with the exponent $\alpha=1/2$, for the whole time range \cite{bray_phase,puri}. 
On the other hand, after $t=10$ the $d=3$ data appear parallel to $\alpha=1/3$. 
For accurate estimation of the exponent for a power-law behavior
it is useful to calculate the
instantaneous exponent \cite{huse_1986}
\begin{equation}\label{eq13} 
 \alpha_i=\frac{d \ln \ell}{d \ln t},
\end{equation}
as well as perform finite size scaling (FSS) analysis \cite{d_landau,m_fisher}.
In Fig. \ref{fig2}(a) we plot $\alpha_i$
as a function of $1/\ell$. Clearly, for $d=3$, the convergence of the data set,
in the limit $\ell=\infty$, is consistent with $\alpha=1/3$, while the $d=2$ data converge to $\alpha=1/2$.
Since the data for large $\ell$ in this figure are noisy, to understand the stability 
of the $d=3$ exponent over long period, we perform the FSS analysis (see Fig. \ref{fig2}(b)). We do not
perform this exercise for $d=2$, since, in this case we have already seen that the data are
consistent with the theoretical expectation,
as established previously \cite{bray_phase,puri}. In fact, from here on, unless otherwise mentioned,
all results are from $d=3$ and $T_f=0$.

\par
In analogy with the critical phenomena \cite{m_fisher}, a finite-size scaling method in the domain growth problems 
can be constructed as \cite{suman,heermann_binder,das_epl}
\begin{equation}\label{eq14}
 \ell(t)=LY(y),
\end{equation}
where the finite-size scaling function $Y$ is independent of the system size 
but depends upon $y$ $(=L^{1/\alpha}/t)$, a dimensionless scaling variable. In the long time 
limit $(y \rightarrow 0)$, when $\ell \simeq L$, $Y$ should be a 
constant. At early time $(y >>0)$, on the other hand, the behavior of 
$Y$ should be such that Eq. (\ref{eq2}) is recovered (since the finite-size effects
in this limit are non-existent). Thus
\begin{equation}\label{eq15}
 Y(y>>0) \sim y^{-\alpha}.
\end{equation}
In the FSS analysis, $\alpha$ is treated as an adjustable parameter. 
For appropriate choice of $\alpha$, in addition to observing the 
behavior in Eq. (\ref{eq15}), data from all different values of 
$L$ should collapse onto a single master curve. In Fig. \ref{fig2}(b) we have used
$\alpha=0.35$. The quality of collapse and the consistency of the power-law decay of the 
scaling function with the above quoted exponent, over several 
decades in $y$, confirm the stability of the value. Thus, it was not 
inappropriate for the previous studies \cite{amar_family,holzer} to conclude that the value of $\alpha$ is $1/3$. 
Nevertheless, given the increase of computational resources over last two decades, it is 
instructive to simulate larger systems over longer periods \cite{corberi}, to check if a crossover 
to the theoretically expected exponent occurs at very late time.
\par
\begin{figure}[htb]
\includegraphics*[width=0.4\textwidth]{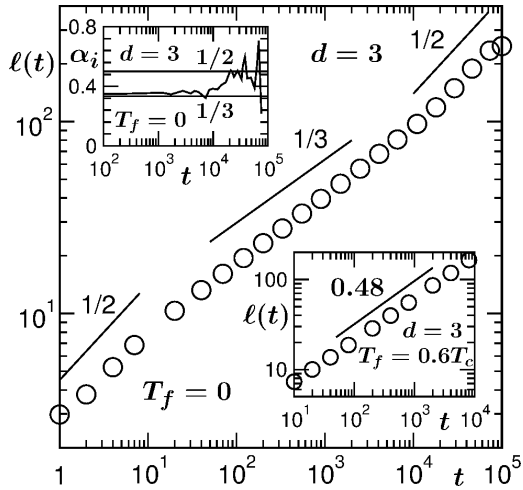}
\caption{\label{fig3}
Log-log plot of $\ell(t)$ vs $t$, for $d = 3$ and $T_f=0$, with 
$L = 512$. The rest of the results are presented for this 
particular system size. The solid lines indicate different
power-law growths, the  exponents being mentioned. The upper inset 
shows instantaneous exponent $\alpha_i$ as a function of $t$, the $x$-axis being in log scale, for the data 
presented in the main frame. The horizontal solid lines there correspond to $\alpha=1/3$ and $1/2$.
The lower inset is same as the main frame but for $T_f=0.6T_c$. The continuous line there
corresponds to a power-law growth with exponent $0.48$.
}
\end{figure}
In Fig. \ref{fig3} we present the $\ell$ vs $t$ data, 
on a log-log scale, from a much larger system size \cite{saikat_pre} 
than the ones considered in Figs. \ref{fig1} and \ref{fig2}. Interestingly, 
three different regimes are clearly visible. A very early  
time regime shows consistency with $\alpha=1/2$.
This is followed by an exponent $1/3$, that stays for about three decades in time. 
Finally, the expected $\alpha=1/2$ behavior is visible, for nearly a decade, 
before the finite size effects appear. In this case, an appropriate 
FSS analysis, to confirm the later time exponent, requires even bigger systems with runs over much 
longer times, which, given the resources available to us, was not possible.
Thus, for an accurate quantification of the asymptotic value of $\alpha$, we restrict ourselves to the analysis via
the instantaneous exponent \cite{huse_1986}. In the upper inset of Fig. \ref{fig3}, we have plotted $\alpha_i$ 
as a function of $t$. The qunatity shows a nice late time 
oscillation around the value $1/2$. This is at variance with the data at high temperature. See the
$\ell$ vs $t$ data, on a log-log scale, from $T_f=0.6T_c$, in the lower inset of Fig. \ref{fig3}. Here
we observe $\alpha=1/2$ for the whole time range. For this data set as well we avoid presenting
results from further analyses.

\par
For $T_f=0$, the crossover that occurs in the time dependence of $\ell$,
may be present in other properties as well \cite{saikat_pre}. These we check next.
For the persistence 
probability, the value of $\theta$ 
was previously estimated \cite{manoj_1}, also from smaller system sizes, to be $\simeq 0.17$. 
\begin{figure}
\includegraphics*[width=0.45\textwidth]{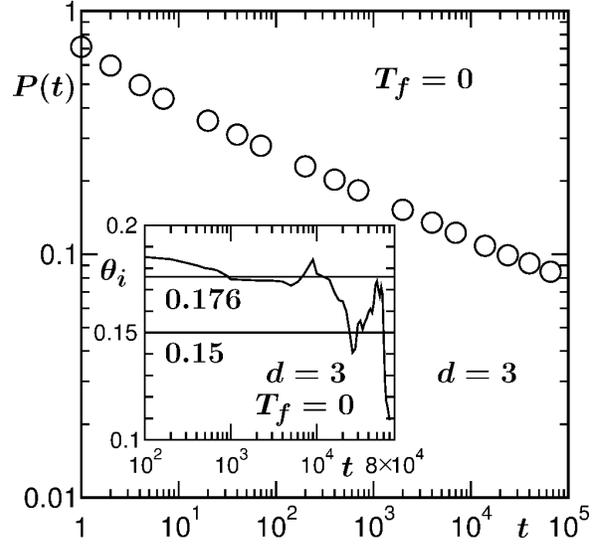}
\caption{\label{fig4}
Log-log plot of the persistence probability, $P(t)$, as a function of $t$, for $d=3$ and $L=512$. 
The inset shows corresponding instantaneous exponent, $\theta_i$, vs $t$, $x$-axis being in a log scale. The 
horizontal solid lines there correspond to the ordinate values $0.176$ and $0.15$. 
The figure has resemblance with Fig. 7(b) of Physical Review E \textbf{93}, 032139 (2016).
These results are from $T_f=0$.
}
\end{figure}
In Fig. \ref{fig4} we show a log-log plot of $P$ vs $t$ and the corresponding 
instantaneous exponent $\theta_i$ (see the inset), calculated as \cite{huse_1986}
\begin{equation}\label{eq16} 
 \theta_{i}=-\frac{d \ln P}{d \ln t},
\end{equation}
vs $t$, for the same (large) system as in Fig. \ref{fig3}. The early time data is consistent with the previous 
estimate. At late time there is a crossover \cite{saikat_pre} to a smaller value 
$\simeq 0.15$, the crossover time being the same as that for the 
average domain size. Here note that, despite improvements \cite{derrida,cueille_jpa,saikat_pre},  
the situation with respect to the calculation
of persistence at non-zero temperature may not be problem free. Thus, for this quantity we avoid presenting
results from $T_f=0.6T_c$.
\par
\begin{figure}
\includegraphics*[width=0.45\textwidth]{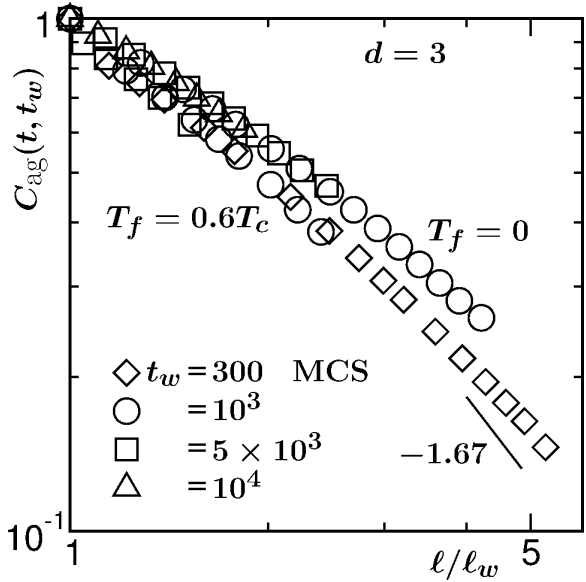}
\caption{\label{fig5}
Log-log plots of the autocorrelation function, $C_{\rm ag}(t,t_w)$, vs $\ell/\ell_w$, for $T_f=0$ and $0.6T_c$.
For each value of $T_f$, results from multiple ages are presented. The solid line corresponds
to a power-law decay, exponent for which is mentioned on the figure.
}
\end{figure}

In Fig. \ref{fig5} we show the plots of $C_{\rm ag}(t,t_{w})$, vs $\ell/\ell_{w}$, from $T_f=0$
and $0.6T_c$, on a
log-log scale, for different values of $t_w$. Good collapse 
of data, for both the values of $T_f$, are visible over the whole range of
the abscissa variable. This, in addition to establishing the
scaling property of Eq. (\ref{eq5}), implies the absence of the finite-size effects \cite{jia_suman,jia_pre}.
On the issue of the finite size effects for the nonconserved Ising model,
a previous study \cite{jia_suman} showed that such effects become important only for $\ell>0.4L$.
The length of our simulations were set in such a way that we are on the edge
of this limit. For $T_f=0$, this can be appreciated from the $\ell$ vs $t$
data in the main frame of Fig. \ref{fig3}.

\par
From a Gaussian auxiliary field ansatz, in the context of the time dependent
Ginzburg-Landau model \cite{bray_phase}, Liu and Mazenko (LM) \cite{liu} constructed a dynamical equation
for $C(\vec{r}_1,t_{w};\vec{r}_2,t)$. For $t\gg t_w$,
from the solution of this equation, they obtained $\lambda\simeq 1.67$ in $d=3$.
The solid line in Fig. \ref{fig5} represents a power-law decay with the above mentioned value
of the exponent. The simulation data, for both values of $T_f$, appear inconsistent with
this exponent. Rather, the simulation results on the log-log scale exhibit continuous bending.
Such bending may be due to the presence of correction to the power law decay at small values of
$x$. Thus, more appropriate analysis is needed to understand these results.

\par
\begin{figure}
\includegraphics*[width=0.45\textwidth]{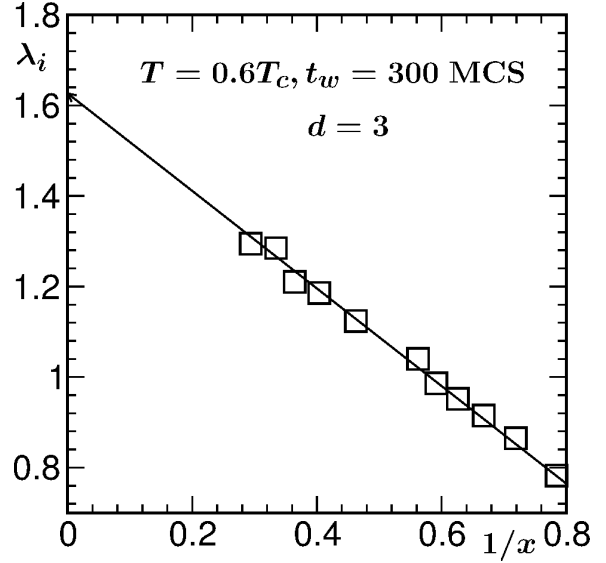}
\caption{\label{fig6}
Plot of the instantaneous exponent, $\lambda_i$, vs $1/x$, for $T_f=0.6T_c$,
with $t_w =300$ MCS. The solid line is a guide
to the eyes.
}
\end{figure}

In Fig. \ref{fig6} we plot the instantaneous exponent \cite{liu,huse_1986,jia_suman}
\begin{equation}\label{eq17}
 \lambda_{i}=-\frac{d \ln C_{\rm ag}}{d \ln x},
\end{equation}
as a function of $1/x$, for $T_f=0.6T_c$. 
A linear behavior is visible, extrapolation of which,
to $x\rightarrow\infty$, leads to $\lambda\simeq 1.62$. The latter
number follows the FH bound \cite{fisher_huse} and is in good agreement with the theoretical prediction
of LM \cite{liu}. Such a linear trend 
provides an empirical full form for the autocorrelation function to be \cite{jia_suman}
\begin{equation}\label{eq18}
 C_{\rm ag}= C_{0}{\exp\bigg(-\frac{B}{x}\bigg)}x^{-\lambda},
\end{equation} 
where $C_0$ and $B$ are constants. A finite-size scaling analysis \cite{jia_suman} using this
expression provided a value even closer to the LM one.
Here note that there already exists \cite{henkel}
a full form, derived from the local scale invariance, for the
decay of $C_{\rm ag}$ during coarsening in the ferromagnetic Ising model.
Validity of this has been demonstrated in studies \cite{lorenz} of $q(>2)$-state Potts model. Even though
derived from a rigorous theoretical method, this expression contains a large
number of unknowns. Thus, analyses of the simulation results by assuming this form is
a tedius process. However, it will be an useful exercise to check how close an approximation 
Eq. (\ref{eq18}) is to this form.

\par
\begin{figure}
\includegraphics*[width=0.45\textwidth]{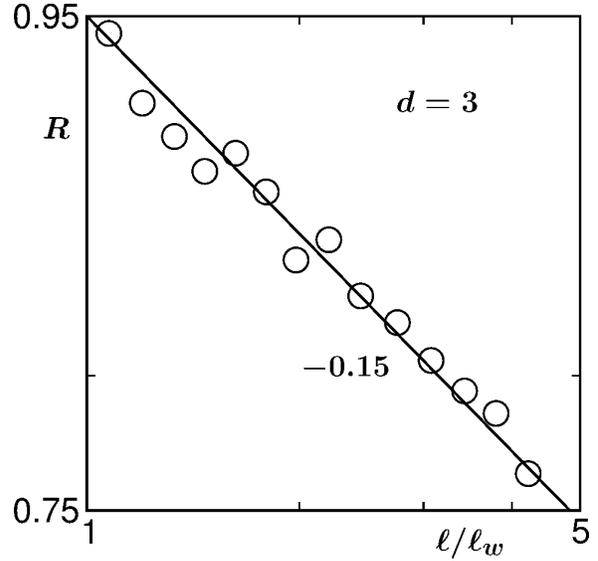}
\caption{\label{fig7}
Log-log plot of the ratio, $R$, between the master curves for the autocorrelations at $T_f=0.6T_c$ and $0$, 
as a function of $x$.
}
\end{figure}

The decay of $C_{\rm ag}$, as seen in Fig. \ref{fig5}, for $T_f=0$ appears slower
than that for $T_f=0.6T_c$. To confirm that, in Fig. \ref{fig7} we plot the 
ratio $R$, between $C_{\rm ag}$ for $T_f=0.6T_c$ and $T_f=0$, on a log-log scale, vs $x$.
Over the whole range of $x$, that covers pre- as well as post-crossover regimes for
domain growth, the data exhibit power-law behavior that can be captured by a single exponent
$\simeq 0.15$. This implies an absence of crossover in the decay of this quantity
and $\lambda \simeq 1.5$, a number significantly smaller than that for $T_f=0.6T_c$.
Whether or not the FH lower bound has actually been violated,
such small value of $\lambda$ calls for further discussion
and calculation of the structural properties.

Yeung, Rao and Desai (YRD) made a more general prediction of the lower bound \cite{yrd},
viz.,
\begin{equation}\label{eq19}
 \lambda \geqslant \frac{d+\beta}{2}, 
\end{equation}
where $\beta$ is the exponent \cite{yeung,lubachev} for the small wave-number ($k$) power-law enhancement of the 
structure factor (the Fourier transform of $C(r,t)$):
\begin{equation}\label{eq20}
 S(k) \sim k^\beta. 
\end{equation}
Here note that $S(k,t)$ has the scaling form (for a self-similar pattern) \cite{bray_phase}
\begin{equation}\label{eq20p}
S(k,t)\equiv \ell^d \tilde{S}(k\ell),
\end{equation}
where $\tilde{S}$ is a time independent master function.
For nonconserved order parameter \cite{yrd}, as in the present case, $\beta=0$. Thus the YRD lower
bound in this case is same as the FH lower bound. The latter bound can as well be appreciated from the 
OJK expression for 
the general correlation function \cite{ojk,puri} in Eq. (\ref{eq4}). This has the form 
\begin{equation}\label{eq21}
C(r;t,t_w)=\frac{2}{\pi}\sin^{-1}\gamma, 
\end{equation}
with 
\begin{equation}\label{eq22}
 \gamma=\bigg(\frac{2\sqrt{tt_{w}}}{t+t_{w}}\bigg)^{d/2}{\rm exp}\bigg[-\frac{r^2}{4D(t+t_{w})} \bigg].
\end{equation}
For $t=t_{w}$, this leads to 
Eq. (\ref{eq8}). For $r=0$ and $t>>t_w$, Eq. (\ref{eq22}) provides
\begin{equation}\label{eq23}
C_{\rm ag}(t,t_{w}) \sim \bigg( \frac{t}{t_{w}}\bigg)^{-d/4}.
\end{equation}
For $\alpha=1/2$, the exponent in Eq. (\ref{eq23}) provides $\lambda=d/2$, which coincides with the FH lower bound. 
Since the latter bound is embedded in Eq. (\ref{eq21}) and the violation of it for $T_f=0$
is a possibility, it is instructive to calculate the structural quantities, viz., $C(r,t)$ and $S(k,t)$,
given that Eq. (\ref{eq21}) contains expressions for the latter quantities as well.
\par
\begin{figure}
\includegraphics*[width=0.45\textwidth]{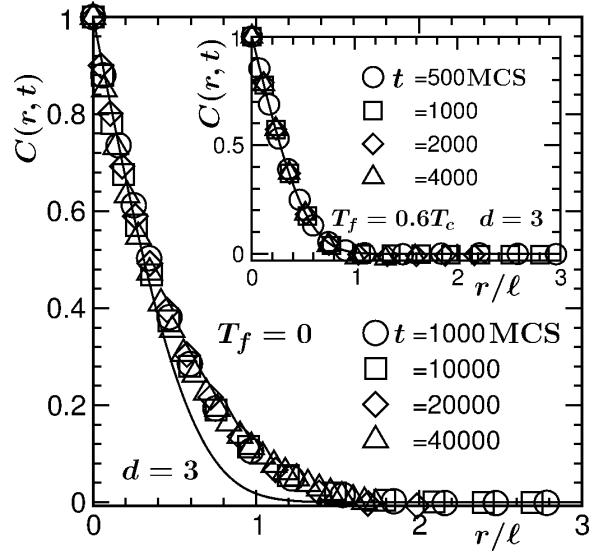}
\caption{\label{fig8}
Scaling plot of the two-point equal time  correlation 
functions from $T_f=0$. The distance along the abscissa has been scaled by the average domain sizes at
different times from which data are presented.
Inset: Same as the main frame, but for $T_f=0.6T_c$. In both the cases the solid curves represent 
the OJK form (see Eq. (\ref{eq8})).
}
\end{figure}
\begin{figure}
\includegraphics*[width=0.45\textwidth]{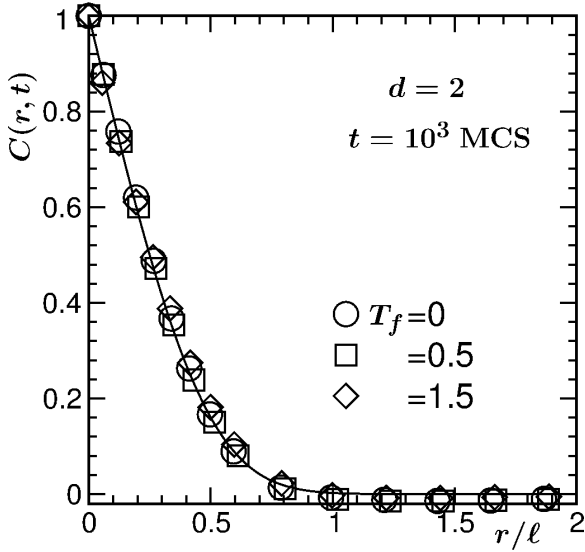}
\caption{\label{fig9}
Scaled correlation functions from different $T_f$ in $d=2$. The continuous curve is the the
OJK function of Eq. (\ref{eq8}).
}
\end{figure}

In Fig. \ref{fig8} we show a scaling plot of $C(r,t)$, vs $r/\ell$, 
for $T_f=0$, $\ell$ being extracted from Eq. (\ref{eq12}).
Nice collapse is visible for data from wide time range. Given that no crossover in $C(r,t)$ is
observed and aging property is strongly related to the structure, it is understandable why a crossover
in the autocorrelation is nonexistent.
The continuous line in this figure is the OJK function \cite{bray_phase,ojk,puri} of 
Eq. (\ref{eq8}). There exists significant discrepancy between the analytical function and the 
simulation results. In the inset 
of this figure we plot the corresponding MC results for $T=0.6T_c$ 
which, on the other hand, shows nice agreement with Eq. (\ref{eq8}). Here note that
in $d=2$ such temperature dependence does not exist \cite{puri}.
For the sake of completeness, this we have demostrated in Fig. \ref{fig9}.
Data from all the temperatures in this case are nicely described by the OJK function. 

\par
\begin{figure}
\includegraphics*[width=0.45\textwidth]{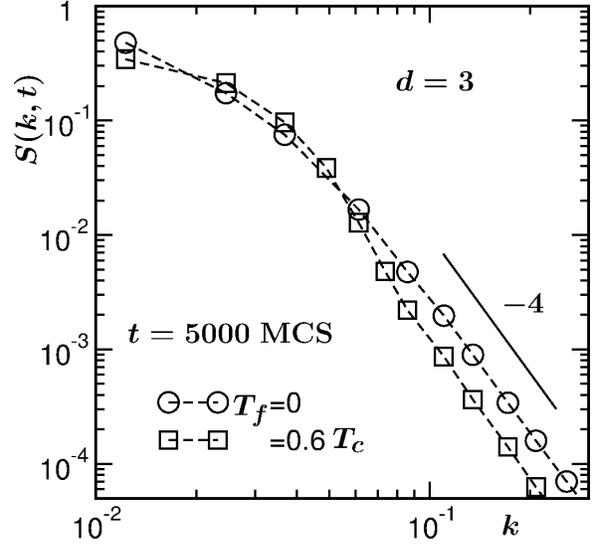}
\caption{\label{fig10}
Plots of the  structure factors, from $T_f=0$ and $0.6T_c$, vs $k$.
The solid line represents the Porod law. For both the temperatures, we have presented results from
$t=5000$. 
}
\end{figure}

In Fig. \ref{fig10} we show a comparison between the structure factors from $T_f=0$ and
$T_f=0.6T_c$, in $d=3$. 
The $k^{-4}$ line in this figure corresponds to
the Porod law \cite{bray_phase,puri,porod} for the long wave-number decay of $S(k,t)$, a consequence
of scattering at sharp interfaces. Data from both the temperatures show reasonable consistency with 
this decay, even in intermediate range of $k$. For the sake of bringing clarity in the small $k$ region,
we did not present the results for the whole range of $k$.
Disagreement between the two cases is visible in the smaller wave-number region.
This may provide explanation for the small value of $\lambda$ for $T_f=0$. For this
purpose, below we provide a further discussion on the derivation of YRD. Starting
from the equal-time structure factors at $t_w$ and $t$, YRD arrived at \cite{yrd}
\begin{equation}\label{eq23p}
C_{\rm ag}(t,t_w) \leq \ell^{d/2}\int_0^{2\pi/\ell} dk k^{d-1} S(k,t_w)\tilde{S}(k\ell).
\end{equation}
To obtain the lower bound, they used the small $k$ form for $S(k,t_w)$, 
as quoted in Eq. (\ref{eq20}). In Fig. \ref{fig10} we see that, compared to $T_f=0.6T_c$, the
the structure factor for $T_f=0$ starts decaying at a
smaller value of $k$, providing an effective negative value for $\beta$.
This is the reason for such a small value of $\lambda$. 

\par
Given that $T_f=0.6T_c$ lies above the
roughening transition temperature $T_{R}$ ($\simeq 0.57T_c$),
possibility exists \cite{corberi} that the observed differences between kinetics at
the two different values of $T_f$ 
may be related to this 
transition \cite{nolden}. 
Our preliminary 
results for other values of $T_f$ ($< T_{R}$) are suggestive of 
this. However, more systematic study is needed. Here note that at the roughening transition
the inteface thickness diverges.
We mention here, most of the previous studies with Glauber \cite{glauber} Ising model 
focused on $d=2$, for which there is no roughening transition.
Nevertheless, question remains, why the crossovers, exhibited by the growth of domains and the decay of persistence,
are missing in the structure and aging? 

\par

\section{Conclusion}

We have studied kinetics of phase transition in $3D$ Ising model via 
the Glauber Monte Carlo simulations \cite{d_landau,glauber}, following quench from $T_i=\infty$ 
to $T_f=0$. Results are presented for domain growth, 
persistence probability, aging and pattern, all of which exhibit 
new features, compared to studies in $d=2$ and quenches to higher 
temperature for $d=3$. The time dependence of the average domain 
size shows consistency with the expected theoretical behavior 
only after an exceptionally long transient period \cite{saikat_pre,corberi}. 
This is reflected in the persistence probability \cite{saikat_pre}.
However, there is no such transient above the roughening transition.

\par
The two-point equal time correlation function does not 
follow the Ohta-Jasnow-Kawasaki form \cite{ojk}, derived for the nonconserved
order-parameter dynamics with scalar order parameter. The latter form, however, 
is found to be consistent with the simulation data above the
roughening transition temperature. Unlike the domain growth and persistence, we 
did not observe any time dependence (crossover) for this observable.
This is reflected in the decay of the autocorrelation function. The latter quantity
appears to have a power-law decay exponent marginally satisfying the Fisher-Huse lower bound.
These results are at deviation with those from $d=2$ for which
there is no roughening transition.

\par
It will be important to understand the
temperature dependence in all these quantities via more systematic studies. This may as well
provide information on the deficiencies in the inputs for the derivation of the OJK function.
One needs to check whether the roughening transition has any role to play with the 
Gaussian approximation, used in the calculation of the structure function.
Furthermore, the reason for long transient in domain growth and persistence deserves attention.
These we aim to address in future works. 
\par
\section*{Acknowledgement}
The authors acknowledge financial support from Department of Science and Technology, India.
SKD is grateful to Marie Curie Actions Plan of European Union (FP7-PEOPLE-2013-IRSES
Grant No. 612707, DIONICOS) for partial support.

* das@jncasr.ac.in

\end{document}